# A polarization maintaining scheme for 1.5 μm polarization entangled photon pair generation in optical fibers

Qiang Zhou, [1,2] Wei Zhang, [1] Tian-zhu Niu, [1] Shuai Dong, [1] Yi-dong Huang, [1] and Jiang-de Peng [1]

[1] Tsinghua National Laboratory for Information Science and Technology, Department of Electronic Engineering, Tsinghua University, Beijing, 100084, P. R. China
[2] betterchou@gmail.com

**Abstract.** In this paper, the generation of polarization entangled photon pairs at 1.5 μm is experimentally demonstrated utilizing a polarization maintaining all-fiber loop, consisting of a piece of commercial polarization maintaining fiber and a polarization beam splitter/combiner with polarization maintaining fiber pigtails. The polarization entangled two photon state of $1/\sqrt{2}\left(|H_s\rangle|H_i\rangle+e^{i\varphi}|V_s\rangle|V_i\rangle\right)$ can be generated while the relative phase $\varphi$ can be adjusted by manipulating the polarization state of the pump light. A quantum state tomography measurement is performed to analyze the entanglement characteristic of the generated quantum state. In the experiment, the polarization entangled Bell state $|\Phi^+\rangle$ is generated with a entanglement fidelity of 0.97±0.03 and a purity of 0.94±0.03 demonstrating that the proposed scheme can realize polarization entangled photon pair generation with polarization maintaining property which is desired in applications of quantum communication and quantum information.

## 1. Introduction

Polarization entanglement at 1.5 μm is a fundamental resource for quantum information technologies, such as quantum key distribution [1,2], quantum communication [3], and quantum information processing [4,5]. Traditionally, polarization entanglement is generated by the spontaneous parametric down conversion processes (SPDC) in nonlinear crystals [6], which has been widely used in experiments of quantum optics. In recent years, with the tendency of developing practical quantum techniques for quantum engineering, various new methods for polarization entanglement generation are proposed and investigated, including the SPDC in period poled crystal waveguides [7,8] and period poled fibers [9], and spontaneous four wave mixing processes (SFWM) in third order nonlinear waveguides, i.e. optical fibers [10-17], and silicon waveguides [18-20]. Among them, the polarization entanglement generation based on SFWM in optical fibers attracts much attention, since it is compatible with the current optical fiber communication technologies, holding the promise to develop practical quantum light source at 1.5 μm band. Several schemes have been proposed to realize the polarization entanglement generation in optical fibers, employing sophisticated optical designs, time-multiplexing scheme [10,11] and the scheme based on polarization diversity loop [12-14], to generate two orthogonally polarized correlated two photon states and make them coherently superposed in time and space. However, in these schemes fine polarization adjustments are needed to minimize the unwanted polarization rotation, which is inevitable in traditional single mode fibers, and sensitive to the temperature and strain variation due to the environment changes. The ways to maintain the performance of these schemes in long term are still under investigation. Recent researches show that optical fibers with birefringence, such as polarization maintaining fibers (PMFs) and some kinds of micro-structured fibers (MSFs), can be utilized to realize polarization entanglement generation without unwanted polarization rotation [15,16]. In these schemes the two orthogonally polarized correlated two photon states are generated along the two fiber polarization axes, respectively. The fiber birefringence change due to the environment variation is still an important source affecting the long term stability of the generated polarization entangled two photon states.

Recently, a modified optical loop scheme was used to realize polarization entangled photon pair generation in silicon wire waveguide, in which only one polarization mode of the silicon wire waveguide was used to avoid the large difference between the correlated photon pair generation efficiencies of the two polarization modes in silicon wire waveguides [18, 19]. Similar experiment design was also used in a recent work of polarization entanglement generation in a piece of PMF at 800 nm band [17]. However, the optical designs of these works need polarization adjustment in fibers or free-space collimation, both of which are undesired in real applications for the inconvenience in developing practical equipment and is sensitive to the environment variation. In this paper, we demonstrate a polarization entanglement generation scheme utilizing a polarization maintaining all-fiber loop, which consists of a polarization beam splitter/combiner with PMF pigtails and a piece of polarization maintaining DSF (PM-DSF). In our proposed scheme, two orthogonally polarized correlated two photon states are generated in the same polarization axis of the PM-DSF, providing a way to solve the problems of unwanted polarization rotation and fiber birefringence change due to the environment variation simultaneously. Furthermore, this scheme realizes stable polarization entanglement generation without any polarization adjustment and free-space collimation, showing a polarization maintaining property desired in applications of quantum communications and quantum information processing.

## 2. The principle of the polarization entangled photon pair generation in the PMAL

Figure 1 shows the principle of the polarization entangled photon pair generation in the polarization maintaining all-fiber loop (PMAL). The polarization beam splitter/combiner (PM-PBS/C) has two polarization axes denoted by $H$ and $V$, while the PMF has two polarization axes denoted by $X$ (slow axis) and $Y$ (fast axis), as shown in the inset of Fig. 1. The PMF pigtails of the PM-PBS/C are spliced with the two ends of the PM-DSF, aligning both the $H$ and $V$ axes of the PM-PBS/C to the $X$ axis of the PM-DSF as shown in Fig. 1. The linearly polarized pulsed pump light is injected into the PMAL through the common end of the PM-PBS/C, with a polarization direction of 45 deg. respecting to the two polarization axes of the PM-PBCS/C. Hence, the $H$ and $V$ components of the pump light would have the same power level. Then the two pump components inject into the PM-DSF in different directions and polarized along the $X$ axis. The pump component propagating in the clockwise (CW) direction generates correlated two photon state of $|X_s\rangle|X_i\rangle_{CW}$, while $|X_s\rangle|X_i\rangle_{CCW}$ is generated by the pump component propagating in counter-clockwise (CCW) direction, where *s* and *i* denote the signal and idler photons. The two generated correlated two photon states inject into the PM-PBS/C through the ports of *H* and *V*, respectively. The temporal and spatial superposition of the two correlated two photon states would be achieved, which may realize the polarization entanglement generation with a state of $1/\sqrt{2}(|H_s\rangle|H_i\rangle+|V_s\rangle|V_i\rangle)$ and output from the common end of the PM-PBS/C. It is worth noting that in the proposed scheme the polarization state change and the polarization mode dispersion (PMD) effect should be eliminated in the PMAL [21], due to the generation and the propagation of the two photon states are along the same polarization principle axis of the PMF. Additionally, the stability of the scheme is determined intrinsically by the reciprocity of the PMAL.

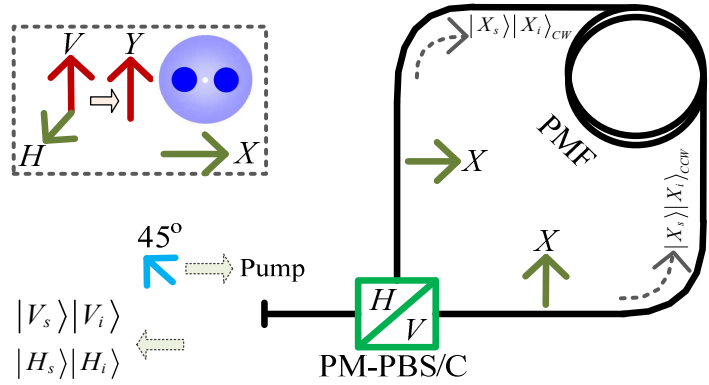

Figure 1. The schematic of the polarization entangled photon pair generation in the polarization maintaining all-fiber loop (PMAL); polarization beam splitter/combiner (PM-PBS/C), polarization beam splitter/combiner with PMF pigtails; PMF, polarization maintaining fiber.

## 3. Experimental demonstration of polarization entangled photon pair generation in the PMAL

We demonstrate the proposed scheme experimentally. Figure 2 shows the experimental setup, which is based on the commercial components at 1.5 μm telecomm band. The pulsed pump light is generated based on the master oscillator power amplifier (MOPA) scheme. The seed pulse is obtained by spectral filtering of the output of a passive mode-locked fiber laser using a density wavelength-division multiplexing (DWDM) device at 1552.52 nm (ITU-C31). The power of the pump is amplified by an erbium doped fiber amplifier (EDFA). The output of the EDFA passes through a filter system consisting of four ITU-C31 DWDM devices, in order to suppress the amplified spontaneous emission of the EDFA and achieve a side-band rejection of >120 dB. The repetition rate and pulse width of the pump light is 3.88 MHz and about ten picoseconds (estimated by a spectral width of 0.52 nm), respectively. A variable optical attenuator (VOA) is used to adjust the power level of the pump light. The polarization state of the pump is controlled by a polarizer (P1), a rotatable quarter-wave (QWP1) plate and a rotatable half-wave plate (HWP1). A three ports circulator with PMF pigtails (PM-C) is used to inject the pump light into the PMAL and collect the output polarization entangled photon pairs from the PMAL. The pulsed pump light inject into the port 1 of the PM-C, and then turns to port 2, which is connected with the PMAL. In the experiment, a piece of PM-PMF (DS15-P S-U40A, fabricated by Fujikura Ltd.) with a length of 50 meters is used in the PMAL. The PM-DSF is cooled by the liquid nitrogen (77 K) in order to reduce the Raman noise [22]. Correlated/entangled photon pairs would be generated in the PM-DSF through SFWM processes. The output of the PMAL input into the port 2 of the PM-C, and output from port 3 of the PM-C finally. The residual pump light and the generated correlated/entangled photon pairs are separated through a filtering and splitting system based on commercial DWDM devices. The central wavelength of signal and idler photons are at 1555.75 nm and 1549.32 nm, i.e., the ITU channels of C27 and C35, respectively. A pump light isolation of >120 dB is achieved at either signal or idler photon wavelength. The -3 dB spectral width of signal and idler side photons are 0.51 nm and 0.52 nm, respectively. The coupling losses of signal and idler photons are -3.3 dB and -3.1 dB, respectively, including the losses of PBS/C, PM-C and the filtering and splitting system. In order to measure the polarization entangled property of the generated photon pairs, two polarization analyzers are added following the filtering and splitting system, as shown in the dashed square in Fig. 2. Each polarization analyzer has a paddle fiber PC, a rotatable QWP, a rotatable HWP and a polarizer (PC1, QWP2, HWP2 and P2 for the signal side; PC2, QWP3, HWP3 and P3 for the idler side). The insertion losses of the polarization analyzers are -0.8 dB and -1.2 dB for signal and idler sides, respectively. The collected signal and idler photons are detected by two commercial single photon detectors SPD1 and SPD2 (ID Quantique, ID201), respectively. The SPDs operate under Geiger mode with 2.5 ns detection window and are triggered with the residual pump light detected by photo-detector (PD). The detection efficiencies of SPD1 and SPD2 are 21.8% and 22.6%, respectively. The dark count rates are $5.82 \times 10^{-5}$ per gate and $4.60 \times 10^{-5}$ per gate, respectively. The output signals of the SPDs are sent into a field programmable gate array (FPGA) based counting system, which records the coincidence and the accidental coincidence counts. In the experiment, the pump power level is monitored by a 50/50 fiber coupler with a power meter (PM) in the residual pump light path of the filtering and splitting system.

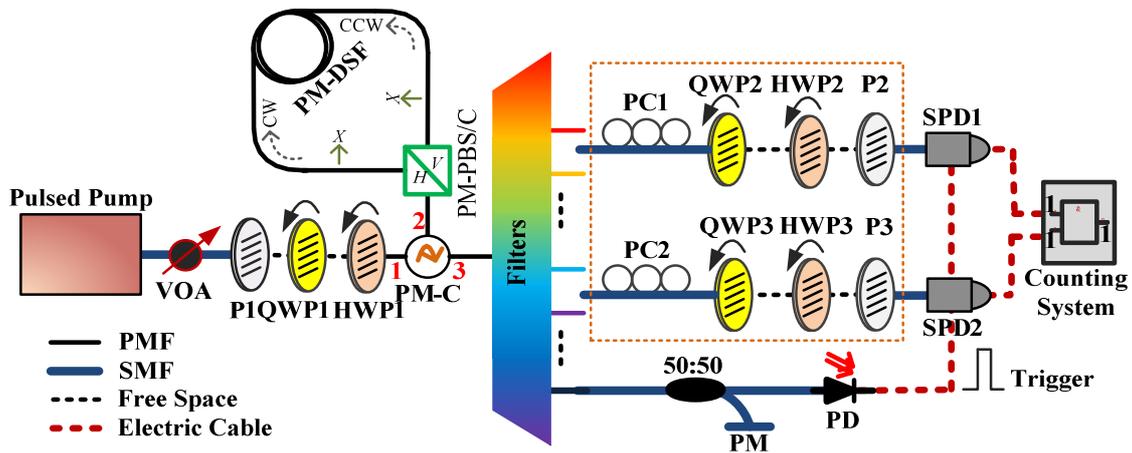

Figure 2. Experiment setup; VOA, variable optical attenuator; P, polarizer; HWP, half-wave plate; PM-C, polarization maintaining circular; PM-PBS/C, polarization maintaining polarization beam splitter and combiner; PM-DSF, polarization maintaining dispersion shifted fiber; PC, polarization controller; QWP, quarter-wave plate; PM, power meter; PD, photon detector; SPD, single photon detector.

Before the polarization entanglement property measurement the initialization of the polarization analyzers needs to be taken. Firstly, the angle of the HWP1 is set to 0 deg., which means that the pump light is polarized along the *V* axis of the PM-PBS/C. Under this condition, the pump light propagates in the PMAL in CCW direction and generates the correlated two photon state of $|H_s\rangle|H_i\rangle$. Then the signal side polarization analyzer can be aligned to *H* axis of the PM-PBS/C by adjusting the PC1 to achieved maximum single side count of the signal photon, while the idler side polarization analyzer can also be aligned to the *H* axis of the PM-PBS/C by adjusting PC2 by the same way.

In the experiment, the generated two photon state should be $1/\sqrt{2}(|H_s\rangle|H_i\rangle + e^{i\varphi}|V_s\rangle|V_i\rangle)$. The phase difference $\varphi$ between $|H_s\rangle|H_i\rangle$ and $|V_s\rangle|V_i\rangle$ is introduced in the experimental setup with $\varphi = 2\varphi_p + \varphi_b$, where $\varphi_b$ is a constant related to the residual birefringence due to the length difference between the pigtailed PMFs at the port 1 and 3 of the PM-C; $\varphi_p$ is determined by the polarization state of the pump light injected in the port 1 of PM-C. Hence the generated state can be adjusted by manipulation the polarization state of pump light [15]. It is worth noting that the PMAL acts as a HWP set to 45°, hence the temporal walk-offs between *H* and *V* polarization components in the PMF pigtails of PM-C compensate itself (port 2) and each other (port 1 and 3). In the experiment setup, PM-C pigtails are coiled in the same way, and the length difference between the PM-C pigtails is controlled less than 1 cm, which can lessen the environment variation impact due to the length difference between the PMF pigtails of PM-C, ensure the temporal and spatial superposition between $|H_s\rangle|H_i\rangle$ and $|V_s\rangle|V_i\rangle$ and increase the stability of $\varphi$ significantly.

The quantum state tomography (QST) measurement is performed to character the density matrix of the generated polarization entangled two photon state [23]. The output density matrix is estimated through maximum likelihood method [24-26]. In the experiment, coincidence count and accidental coincidence count are measured under sixteen settings of the two polarization analyzers at the signal and idler sides, which project the detected signal and idler photons onto different combinations of the polarization bases for detection, including *H*, *V*, $D = (H+V)/\sqrt{2}$, $L = (H-iV)/\sqrt{2}$ and $R = (H+iV)/\sqrt{2}$. The difference between the coincidence count and the accidental coincidence count is that the coincidence count of signal and idler photons are generated from the same pump pulse or different pump pulses. In the experiment, each result of the coincidence and accidental coincidence count is an average of 5 times measurement with a counting time of 10 seconds. Finally, the density matrix of the generated two photon state can be calculated by a maximum likelihood tomography algorithm utilizing the experimental results of coincidence count under the sixteen polarization analyzer settings.

## 4. Experimental results and discussion

In the experiment, correlated/entangled photon pairs are generated through SFWM while pump pulses propagate in the PM-DSF. The photon pair generation rate quadratically increases with the pump power level [27], while the probability of multi-photon pair event would increase with the photon pair generation rate also. In order to reduce the multi-photon pair events, the photon pair generation rate is set to about 0.01 per pulse in the QST measurement, under an average pump power of 1.58 μW injecting into the PMAL. The average coincidence and accidental coincidence counts are about 90 and 5 per second, respectively, under $|HH\rangle$ or $|VV\rangle$ measurement basis set. The coupling ratios are 46.9% and 45.1% for idler and signal sides, respectively, which are agree with the measured coupling losses (-3.3 dB and -3.1 dB for signal and idler, respectively). The difference between the coupling ratios and the coupling losses may be due to the slight spectral mismatch between the pump filters and the signal and idler filters.

For a two photon state of $1/\sqrt{2}(|H_s\rangle|H_i\rangle + e^{i\varphi}|V_s\rangle|V_i\rangle)$, the density matrix $\rho$ can be written as,

$$\rho = \frac{1}{2} \times \left( |H_s\rangle|H_i\rangle\langle H_i|\langle H_s| + e^{-i\varphi}|H_s\rangle|H_i\rangle\langle V_i|\langle V_s| \right. \\ \left. + e^{i\varphi}|V_s\rangle|V_i\rangle\langle H_i|\langle H_s| + |V_s\rangle|V_i\rangle\langle V_i|\langle V_s| \right) \quad (1)$$

The absolute values of off-diagonal terms of $\rho$ is due to the superposition of the photon pair wave-function, represent a signature of entanglement and the state purity, which vanishes if the two photon state is classically correlated light, while 0.5 represents a entangled state with high state purity.

Firstly, polarization entangled two photon states are generated by injecting a linearly polarized pump light into the PMAL with a polarization direction of 45 deg. respecting to H/V basis. The reconstructed density matrix $\rho$ of the generated two photon state is shown in Fig. 3 (a) and (b), the real and imaginary parts of $\rho$ respectively, which is represented using $|HH\rangle$, $|HV\rangle$, $|VH\rangle$ and $|VV\rangle$ as basis set. This result is obtained using the true coincidence counts, in which the contribution of accidental coincidence count is subtracted. The experimental results in Fig. 3 (a) and (b) suggest that the measured two photon state is $|\phi\rangle = 1/\sqrt{2}(|H_s\rangle|H_i\rangle + e^{i(0.24\pm0.01)}|V_s\rangle|V_i\rangle)$, with a residual phase of 0.24 rad. due to the residual birefringence in the experiment setups. The fidelity of the measured entangled photon pairs is 0.95±0.04 calculated by $F = \sqrt{\langle\phi|\rho|\phi\rangle}$; the purity of the output state is 0.88±0.04, both of which show a clear characteristic of entanglement. The raw fidelity is 0.85±0.03 without subtracting the accidental coincidence counts, which also exceeds the maximum limit of classically correlated light (0.5).

Then the residual phase is compensated by setting the pump light into elliptically polarized with a polarization extinction ratio of 9.2 dB and its long elliptical axis orienting at 45 deg. respecting to H/V basis. Fig. 3 (c) and (d) show the reconstructed density matrix $\rho'$ of the generated two photon state. The experimental results in Fig. 3 (c) and (d) suggest that the measured two photon state is the polarization entangled Bell state $|\Phi^+\rangle$ with a fidelity of 0.97±0.03 calculated by $F = \sqrt{\langle\Phi^+|\rho'|\Phi^+\rangle}$, while the raw fidelity is 0.88±0.03 without subtracting the accidental coincidence counts. The purity of the output state is 0.94±0.03. The experimental results suggest that the residual phase difference due to the residual birefringence is compensated by manipulating the polarization state of the pump light.

The experimental results shown in Fig. 3 demonstrate that polarization entangled two photon state are successfully generated based on the polarization maintaining all-fiber loop scheme. The imperfections of the measured entangled two photon state are mainly due to the polarization dependent loss and the angle error of the manual rotatable QWPs and HWPs used in polarization analyzers, in which the angle error is ±3° in the experiment. It is worth noting that the filtering and splitting system can be substituted by commercial DWDMs with polarization maintaining property and with polarization walk-off effect between $|H_s\rangle|H_i\rangle$ and $|V_s\rangle|V_i\rangle$ components compensated. Hence, the demonstrated scheme has the potential on generating and measuring the polarization entanglement with polarization maintaining property which would be desired in quantum information engineering.

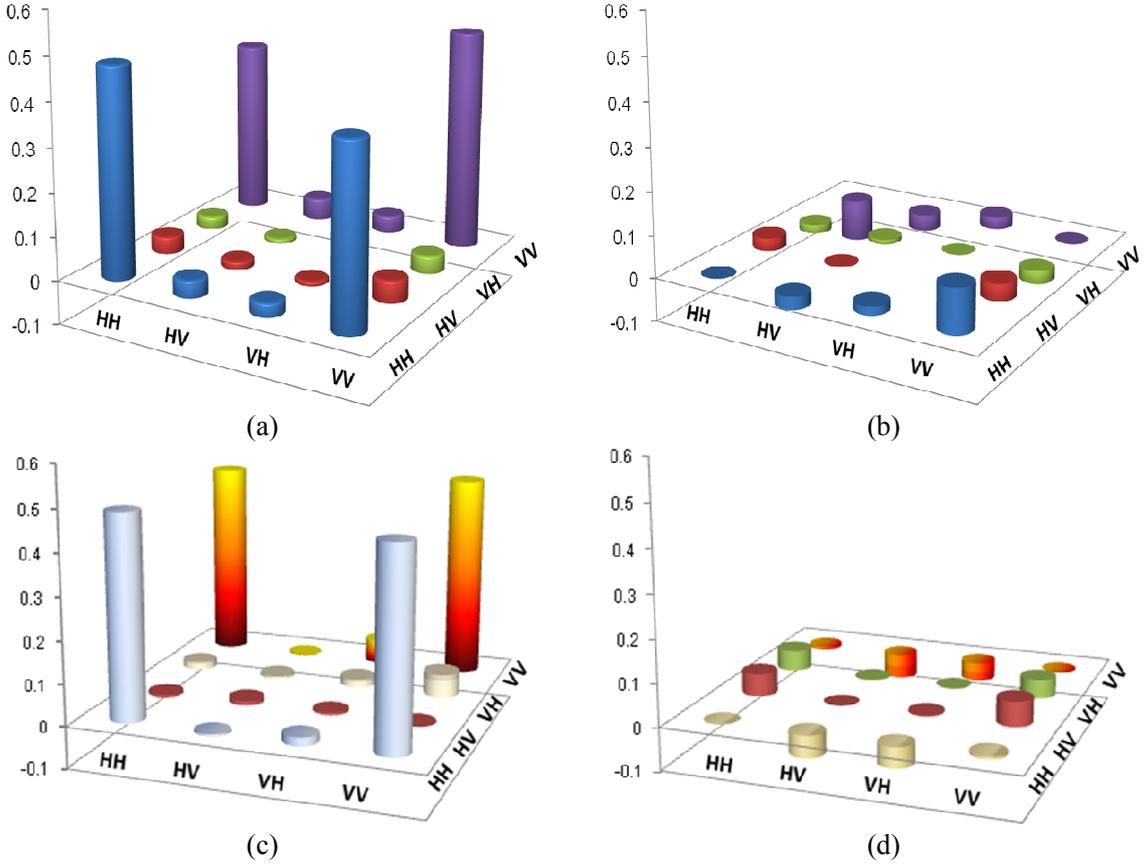

Figure 3. Experiment results of the reconstructed density matrix. (a) the real parts, and (b) the imaginary parts of the reconstructed density matrix of two photon state generated with a linearly polarized pump light; (c) the real parts, and (d) the imaginary parts of the reconstructed density matrix of two photon state generated with a elliptically polarized pump light.

## 5. Conclusions

In conclusion, a PMAL based scheme for 1.5 μm polarization entangled photon pair generation is proposed and demonstrated experimentally. The PMAL consists of a PM-PBS/C and a piece of PM-DSF. The polarization entangled two photon state of $1/\sqrt{2}(|H_s\rangle|H_i\rangle+e^{i\varphi}|V_s\rangle|V_i\rangle)$ can be generated while the relative phase $\varphi$ can be adjusted by manipulating the polarization state of the pump light. A quantum state tomography measurement is performed to analyze the entanglement characteristic of the generated quantum state. In the experiment, the polarization entangled Bell state $|\Phi^+\rangle$ is generated with a entanglement fidelity of 0.97±0.03 and a purity of 0.94±0.03. The polarization maintaining design of our scheme eliminates the unwanted polarization rotation in single mode fibers, while the reciprocity in the PMAL ensures the stability of the phase difference between the two correlated two photon states of the polarization entangled photon pairs. Hence, it provides a way to realize high performance polarization entangled photon pair generation with polarization maintaining property showing a great potential on realizing practical quantum light sources.

## 6. Acknowledgement

This work is supported in part by 973 Programs of China under Contract No. 2011CBA00303 and 2010CB327606, Tsinghua University Initiative Scientific Research Program, Basic Research Foundation of Tsinghua National Laboratory for Information Science and Technology (TNList). Qiang Zhou acknowledges the support from China Postdoctoral Science Foundation.